\documentclass[aps,pra,showpacs,preprintnumbers,amsmath,amssymb]{revtex4}
\usepackage{graphicx}
\usepackage{dcolumn}
\usepackage{bbm}

\begin{document}
\bibliographystyle{apsrev}

\preprint{APS/123-QED}

\title{Optimal control of magnetization dynamics in ferromagnetic heterostructures by spin-polarized currents
}

\author{M. Wenin}
\email{Markus.Wenin@uni-graz.at}
\author{A. Windisch}
\email{06windis@edu.uni-graz.at}

\author{W. P\"otz}%
 \email{Walter.Poetz@uni-graz.at}
\affiliation{%
Institut f\"ur Physik, Theory Division Karl Franzens Universit\"at
Graz,\\ Universit\"atsplatz 5, 8010 Graz, Austria
}%
\date{\today}

\begin{abstract}
We study the switching-process of the magnetization in a
ferromagnetic-normal-metal multilayer system by a spin polarized
electrical current via the spin transfer torque. We use a spin
drift-diffusion equation (SDDE) and the Landau-Lifshitz-Gilbert
equation (LLGE) to capture the coupled dynamics of the spin
density and the magnetization dynamic of the heterostructure.
Deriving a fully analytic solution of the stationary SDDE
we obtain an accurate, robust, and fast self-consistent model for
the spin-distribution and spin transfer torque inside general
ferromagnetic/normal metal heterostructures. Using optimal control
theory we explore the switching and back-switching process of the
analyzer magnetization in a seven-layer system. Starting from a
Gaussian, we identify a unified current pulse profile which
accomplishes both processes within a specified switching time.

\end{abstract}


\maketitle
\section{\label{Int} Introduction}
Spin transfer torque in nanoscaled ferromagnetic/normal--metal
(FN) heterostructures has potential application for data storage
and manipulation \cite{Kisev,Dassow,Poetz}. Apart from the
experimental studies many theoretical investigations have been
made since the pioneering work by Slonczewski and Berger
\cite{Slonczewski,Berger,Wilczynsky,Alpakov,Berkov}. The problem
to describe the physics in FN heterostructures arises from the
need to consider the dynamics of the conduction electrons as spin
carriers and the dynamics of the localized magnetic moments in
parallel and in different regions of the heterostructure. The
electron dynamics is faster by several orders of magnitude than
that of the latter \cite{Li2}.  Moreover, the spin dynamics in
normal metal regions differs significantly from that in
ferromagnetic regions:  the former is characterized by fast
diffusion and slow spin relaxation, while in the latter the
opposite is the case. This time hierarchies make it difficult to
provide a fully numerical solution. A Boltzmann--transport theory
for magnetic multilayer systems including the spin was developed
by Valet--Fert \cite{Valet,Zhang}. On the next level of
approximation a drift--diffusion equation was applied for mobile
spins \cite{Barnas}. The dynamics of the localized magnetic
moments is governed by the Landau--Lifshitz--Gilbert equation
(LLGE), extended by additional spin transfer terms. A similar
investigation has been performed for semiconductor/ferromagnetic
multilayers assuming ballistic transport, but using
non--equilibrium Green's functions \cite{Datta}.

In this paper we utilize this time hierarchy and base our model on
an exact stationary solution to the spin drift--diffusion equation
which we were able to obtain for constant electric current and
arbitrary but piece-wise constant layer parameters. We solve
self--consistently the LLGE and the spin drift--diffusion equation
(SDDE) for the conduction electrons in an external magnetic field
to explore switching scenarios as a function of current pulse
profiles.

The paper is organized as follows. In Sec.~\ref{sec1} we present
our model for FN multilayer system. Sec.~\ref{sec2} and
Sec.~\ref{sec3} are devoted to the mathematical description of the
magnetization dynamics (LLGE) and the dynamics of the conduction
electrons (SDDE) respectively.  The exact solution of our SDDE is
given, with details deferred to the Appendix. In Sec.~\ref{sec4}
we present numerical results for a symmetric seven--layer system.
Optimal current pulse profiles to switch the magnetization in a
given time from parallel to antiparallel state (and in opposite
direction) is shown.  Our results are compared with our fully
numerical simulations to confirm the validity of our approach. Our
results regarding critical switching currents versus switching
time agree well with earlier work by others\cite{Fuchs2,Meng} .
\section{\label{sec1}Model}
Our model of the heterostructure assumes three different physical
building blocks: (i) the normal--metal leads and spacer layers,
(ii) ferromagnetic polarizers , and (iii) ferromagnetic analyzers.
The leads and the spacer layers are chosen to be nonmagnetic (N)
metals with equal material properties. A lead is assumed to be
infinitely thick and serving as a spin bath with vanishing spin
polarization. We describe a wide ferromagnetic hard polarizer
layer ($P_1$, $P_2$ in Fig.~\ref{G0}) as static and homogeneous. A
thin ferromagnetic (soft) analyzer layer (region A in
Fig.~\ref{G0}) is treated as a ferromagnetic mono--domain
described by a single time--dependent variable, a unit--vector
$\textbf{m}(t)$ pointing in the direction of the magnetization
\cite{Stiles,Tserkovnyak}. The conduction spin--electrons are
treated as classical magnetic moments moving in an external
magnetic field created by localized magnetic moments in the
ferromagnet. The spin density $\textbf{S}(\textbf{x},t)$ is the
dynamical variable to describe the spin distribution
\cite{Jaroslav2}. It is defined for an isolated ferromagnet with
magnetization direction $\textbf{m}$ as
\begin{equation}
\textbf{S}=n\mathcal{P}\frac{\hbar}{2}\textbf{m}~.
\end{equation}
Here $n=n_{\uparrow}+n_\downarrow$ is the free electron number
density, where $n_{\uparrow,\downarrow}$ is the particle density
with spin up/down respectively and
$\mathcal{P}=\frac{n_{\uparrow}-n_\downarrow}{n}$ corresponds to
the spin density polarization, extracted from the experiment
\cite{Ziese}. In this work we use for the spin density the
dimensionless quantity $\textbf{s}=\textbf{S}/n\frac{\hbar}{2}$.
For simplicity we do not consider spin--resolved quantities but
use mean values instead (diffusion constant, electric
conductivity, spin diffusion length etc.).
\section{\label{sec2} Magnetization dynamics}
\subsection{Landau-Lifshitz-Gilbert equation}
The temporal evolution of the magnetization $\textbf{M}$ is
governed by the LLGE
\cite{Landau,Stiles}.
Using the saturation magnetization $M_s$, we
define the quantities $\textbf{M}=M_s\textbf{m}$,
$\textbf{h}=\gamma \textbf{H}$, where $\gamma$ is the gyromagnetic ratio and
$\left(\frac{\partial\textbf{m}}{\partial
t}\right)_{st}=\frac{1}{M_s}\left(\frac{\partial\textbf{M}}{\partial
t}\right)_{st}$ to obtain an equation of motion for the
dimensionless magnetization:
\begin{equation}\label{LLG}
\frac{d\textbf{m}}{dt}=-\frac{1}{1+\alpha^{2}}\textbf{m}\times\textbf{h}-
\frac{\alpha}{1+\alpha^2}\textbf{m}\times(\textbf{m}\times\textbf{h})+\left(\frac{\partial\textbf{m}}{\partial
t}\right)_{st}~.
\end{equation}
Here $\textbf{h}=\textbf{h}_{an}+\textbf{h}_{ex}$
is the effective field containing the anisotropy field and
external fields measured in units of a frequency and $\alpha$ the Gilbert damping constant. With
a unit vector $\textbf{n}$ we set for the anisotropy field
\begin{equation}\label{han}
\textbf{h}_{an}=\omega_{an}\textbf{n}(\textbf{m}\cdot\textbf{n})~,
\end{equation}
where $\omega_{an}$ is the corresponding frequency. $\left(\frac{\partial\textbf{M}}{\partial t}\right)_{st}$ denotes
the spin-transfer term \cite{Slonczewski,Berger,Tserkovnyak},
\begin{equation}\label{T}
\left(\frac{\partial\textbf{m}}{\partial
t}\right)_{st}=\xi\textbf{m}\times(\Delta\textbf{I}_s\times\textbf{m})~.
\end{equation}
Here
$\Delta\textbf{I}_s\equiv(\textbf{I}_s)_{in}-(\textbf{I}_s)_{out}$
stands for the spin current absorbed inside the domain, whereas
$\xi$ is a constant \cite{Stiles1,Li}. Without external torque the
equilibrium magnetization is either parallel (P) or antiparallel
(AP) to $\textbf{n}$.

\subsection{Dipole field}
In this paper we consider the control of the magnetization by spin
currents only. So the only contribution to $\textbf{h}_{ex}$ from
the outside are the dipole fields originating from the polarizers.
In order to obtain a simple result and a crude estimate of the
order of magnitude of the dipole fields we consider a polarizer
(here written for $P_1$ in Fig.~\ref{G0}) as a cylinder with
radius $R$ and thickness $x_1$ which is homogeneous magnetized and
compute the field at the position $x_m=(x_2+x_3)/2$. Evaluation of
the general integral for a dipole density \cite{Jackson} we obtain
($\{\textbf{e}_x,\textbf{e}_y,\textbf{e}_z\}$ is the
canonical basis)
\begin{equation}\label{Hdd}
\textbf{H}_{d-d}=\frac{1}{4}M_s\textbf{e}_z~\left\{\frac{x_m-x_1}{\sqrt{R^2+(x_1-x_m)^2}}-\frac{x_m}{\sqrt{R^2+x_m^2}}\right\}.
\end{equation}
\section{\label{sec3}Dynamics of the conduction electrons}
A detailed derivation of the balance equation for the spin density
$s_j(\textbf{x},t)$ is a many particle problem \cite{Heide}.
We use the phenomenological expression for the spin current
density \cite{Jaroslav2},
\begin{equation}\label{phenomcurr}
\textbf{j}_k(\textbf{x},t)=-\mu
s_k(\textbf{x},t)\textbf{E}(t)-D(\textbf{x})\nabla \delta
s_k(\textbf{x},t)~.
\end{equation}
$\textbf{j}_k$ is the spin current density for electrons with
spin--polarization along the $k-$axis. $\mu$ is the electron
mobility, which we assume as material--independent, and
$\textbf{E}(t)$ is the time--dependent electric field.
$D(\textbf{x})$ stands for the material--dependent diffusion
constant and $\delta s_k(\textbf{x},t)\equiv
s_k(\textbf{x},t)-s_k^{eq}(\textbf{x},t)$ is the non--equilibrium
spin density (spin--accumulation), $\textbf{s}^{eq}(\textbf{x},t)$
is the space-- and time--dependent equilibrium spin density.  We
compute the latter using the SDDE, as explained in the next
section. Eq.~\eqref{phenomcurr} is in general valid for
ferromagnetic as well as for nonmagnetic materials. Because of
$\nabla\cdot\textbf{E}=0$ inside the metal, we obtain the spin
drift--diffusion equation (SDDE) \cite{Jaroslav2,Zutic},
\begin{eqnarray}\label{SDE}
\frac{\partial \textbf{s}}{\partial
t}=\frac{|e|}{m}\textbf{s}\times \textbf{B}+(\nabla D\cdot
\nabla)\delta\textbf{s}+D \triangle
\delta\textbf{s}+{}\nonumber\\
\mu(\textbf{E}\cdot\nabla) \textbf{s}+\left(\frac{\partial
\textbf{s}}{\partial t}\right)_{sf}~.
\end{eqnarray}
$|e|$ is the elementary charge and $m$ the electron mass ($\mu_0$
is the permeability of the vacuum).
For the spin flip term we make a spin--relaxation--time ansatz,
\begin{equation}
\left(\frac{\partial \textbf{s}}{\partial
t}\right)_{sf}=-\frac{\delta\textbf{s}}{\tau(\textbf{x})}~,
\end{equation}
with the space dependent relaxation time $\tau(\textbf{x})$. For
simplicity we assume an isotropic $\tau$ inside each layer. In
Eq.~\eqref{SDE} the magnetic induction is related to the
magnetization and an external field trough
$\textbf{B}(\textbf{x},t)=\mu_0(\textbf{H}(\textbf{x},t)+\textbf{M}(\textbf{x},t))$.
\subsection{Spin and charge currents}
From now on we will consider quasi--one--dimensional systems along
the $x$--axis, such as sketched in Fig.~\ref{G0}. Using
Eq.~\eqref{phenomcurr} we obtain for the spin current
$\textbf{I}_s$
\begin{equation}
\textbf{I}_s(x,t)=-A\left(\mu E(t)~\textbf{s}(x,t)+D(x)~
\frac{\partial \delta\textbf{s}}{\partial x}(x,t)\right)~.
\end{equation}
Here $A$ is the cross--section of the sample, and
$\textbf{E}(t)=E(t)\textbf{e}_x$. In the drift--diffusion model
the charge current density is given
by \cite{Seeger}
\begin{equation}
j(x,t)=n|e|\mu E(t)-D(x)\frac{\partial n}{\partial x}~.
\end{equation}
We assume homogeneity, $\frac{\partial n}{\partial x}=0$ to obtain
\begin{equation}
j(t)=n|e|\mu E(t)~.
\end{equation}
A related quantity is the drift--velocity $v_d$, used for
numerical computations, defined by $v_d(t)=-j(t)/n|e|$. For
electrons $j$ and $v_d$ have opposite sign. $v_d>0$, $j<0$ means
electrons (spin--carrier) move in the positive $x-$ direction.
\subsection{Equilibrium spin density}
Because we consider an arbitrary movement of the analyzer
magnetization vector $\textbf{m}(t)$ we have a time--dependent
equilibrium spin density (we neglect spin pumping processes
induced by moving magnetization \cite{Tserkovnyak,Taniguchi}). For
a fixed time $t$ the equilibrium spin density
$\textbf{s}^{eq}(x,t)$ is space--dependent, with a first order
approximation (isolated layers),
\begin{equation}\label{stilde}
\textbf{s}^{eq}(x,t)\Big|_{(1)}\equiv\tilde{\textbf{s}}=\left\{%
\begin{array}{ll}
  \mathcal{P}\frac{\textbf{B}(t)}{\mid\textbf{B}(t)\mid}, & \hbox{$x\in F$;} \\
    \textbf{0}, & \hbox{$x\in N$.} \\
\end{array}%
\right.
\end{equation}
This expression reflects our choice of dimensionless spin density
$\textbf{s}$ in Eq.~\eqref{phenomcurr} and Eq.~\eqref{SDE}. To
obtain the equilibrium spin density in the complete structure we
use the general stationary solution Eq.~\eqref{sol1} presented in
the next section, in which we use the first order expression
Eq.~\eqref{stilde} for $\textbf{s}^{eq}(x,t)$. We use the boundary
conditions for transparent interfaces in F/N--junctions
\cite{Jaroslav}: $\textbf{s}(x,t)$ and $\textbf{I}_s(x,t)$ are
continuous. For $E(t)=0$ we obtain in second order
$\textbf{s}^{eq}(x,t)\mid_{(2)}$. In the following we omit the
subscript $\mid_{(2)}$. Note that, for non--collinear magnetic
layers, all components of $\textbf{s}^{eq}(x,t)\neq 0$ in general.
\subsection{Stationary solution of the SDDE for constant
parameters}\label{SDDstatsol}
For a given layer, we consider Eq.~\eqref{SDE} for constant current, constant
material parameters and time-- and space--independent magnetic
field. We use $\textbf{s}^{eq}=\tilde{\textbf{s}}$ given by
Eq.~\eqref{stilde}. Setting $\partial \textbf{s}/\partial t=0$ in
Eq.~\eqref{SDE}, we have for a one dimensional structure the
equation
\begin{eqnarray}\label{SDEstat}
D~\textbf{s}''(x)+ \mu E~
\textbf{s}'(x)+\omega~\textbf{s}(x)\times \textbf{b}_1
-\frac{\textbf{s}(x)-\tilde{\textbf{s}}}{\tau}=0~.
\end{eqnarray}
Here we have defined $\textbf{b}_1=\textbf{B}/|\textbf{B}|$, and
$\textbf{B}=(m/|e|)\omega\textbf{b}_1$, with $\omega$ the Larmor
frequency. The general solution of Eq.~\eqref{SDEstat}
is a quite lengthy expression, containing 6 integration constants,
denoted as $c_1 \ldots c_6$. To find it we split $\textbf{s}(x)$ into two
parts, one part parallel to the magnetic field, and the other
perpendicular to it,
\begin{equation}\label{sol1}
\textbf{s}(x)=\textbf{s}_\parallel(x)+\textbf{s}_\perp(x)~.
\end{equation}
We define an orthonormal, positive oriented basis
$\{\textbf{b}_1,\textbf{b}_2,\textbf{b}_3\}$. One finds for the
parallel part (where $l_d=\mu E \tau$ is the drift length with sign determined by $E$),
\begin{eqnarray}\label{sparal}
\textbf{s}_\parallel(x)=\textbf{b}_1\Bigg\{c_1
\exp\Big[-\frac{x[l_d+\sqrt{l_d^2+4\lambda^2}]}{2\lambda^2}\Big]+{}\nonumber\\
c_2\exp\Big[-\frac{x[l_d-\sqrt{l_d^2+4\lambda^2}]}{2\lambda^2}\Big]\Bigg\}+\tilde{\textbf{s}}~.
\end{eqnarray}
 $\textbf{s}_\parallel(x)$ does not depend either on $|\textbf{B}(t)|$
or the saturation magnetization. The second part is given by
\begin{eqnarray}\label{ssenk}
\textbf{s}_\perp(x)=\textbf{b}_2\Big\{c_3 G_4(x)+c_4
G_3(x)+c_5 G_2(x)+{}\nonumber\\
c_6 G_1(x) \Big\}+ \textbf{b}_3\Big\{c_3 G_3(x)-c_4
G_4(x)-\nonumber\\
c_5 G_1(x)+c_6 G_2(x) \Big\}~.
\end{eqnarray}
Here the functions $G_i(x)$, $i=1,...4$ are given in Appendix~\ref{AppA}.
They depend on the magnetic field and the electric current,
not indicated here to simplify the notation. Eq.~\eqref{sol1} with
Eq.~\eqref{sparal} and Eq.~\eqref{ssenk} present the complete
solution of Eq.~\eqref{SDEstat} used in our numerical simulations.
We make the following remarks:\\
(i) The solution of of the SDDE for spin--orientation--dependent material parameters is straightforward.\\
(ii) Using this solution one can study different boundary
conditions when linking layers.\\
(iii) Because the solutions for spin densities parallel and normal (to the magnetic field)
can be separated, it is immediately possible to refine the
model using different times $\tau_1$ and $\tau_2$ for spin relaxation and dephasing.
\subsection{Validity of the quasi--static solution}\label{Fulltimedependent}
\begin{figure}
\begin{center}
\includegraphics[height=90 mm, angle=-90]{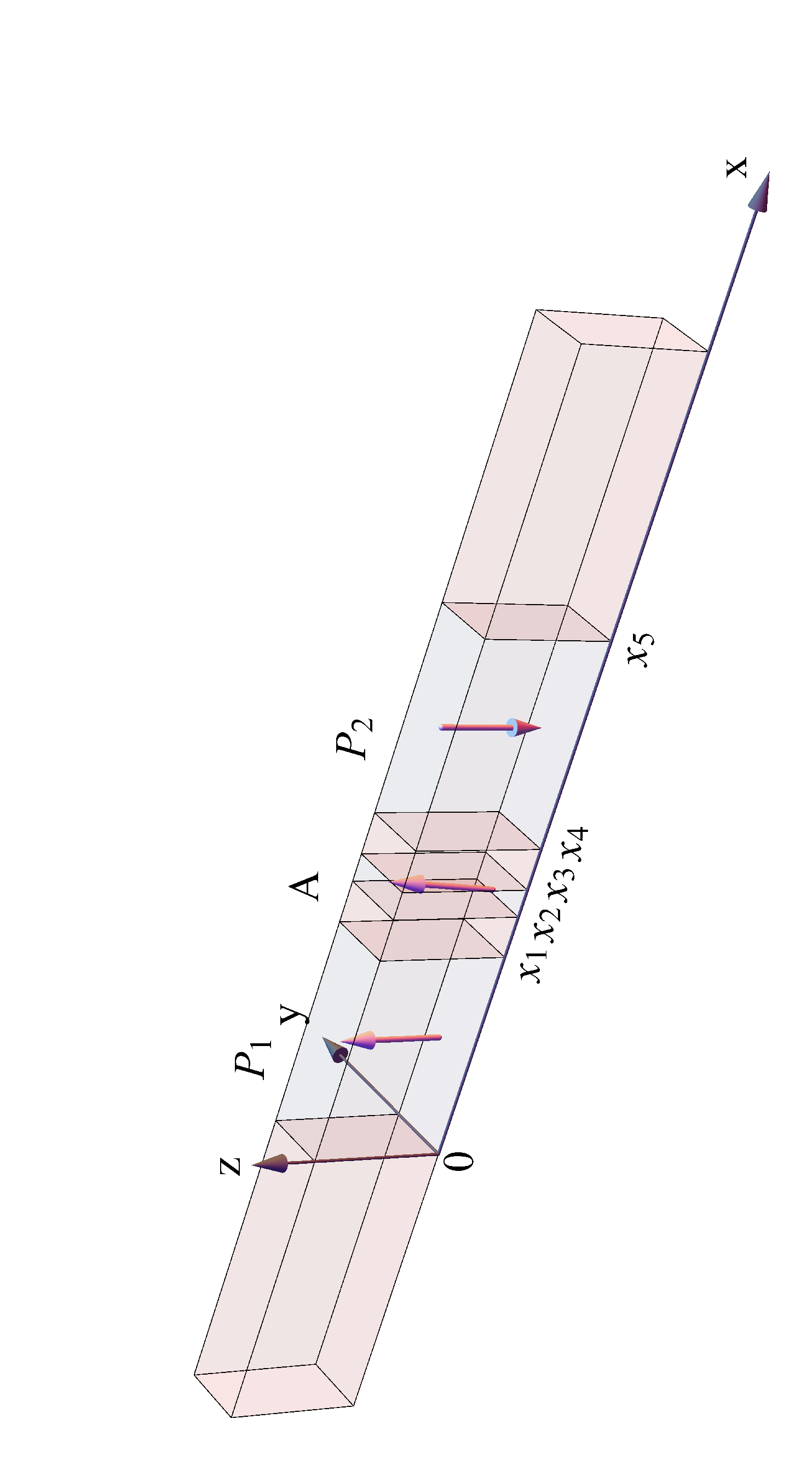}
\caption{\label{G0}(color online) Geometry of the seven--layer
system. The outside layers act as spin--carrier (electron)
reservoirs  with polarization $\mathcal{P}=0$. The regions $P_1$
and $P_2$ are the two polarizers, and A is the analyzer layer
whose magnetization is to be manipulated.}
\end{center}
\end{figure}

Here we develop a scheme to estimate the errors from our
quasi-static approach. We use the stationary solution from the
previous section to compute the spin density for a time--dependent
current and magnetization vector. In general this approximation is
valid as long as the variation of $j(t)$ and $\textbf{m}(t)$ is
slow compared to the shortest relaxation time $\tau$
(quasi--static time--evolution, QSE). A more rigorous estimate of
the accuracy of the QSE in comparison with the solution of the
full time--dependent equation is a non--trivial task. This is due
the different relevant processes and time--scales in the different
layers. To get a quantitative picture we set
$\textbf{s}(x,t)=\textbf{s}_{qs}(x,t)+\delta\textbf{s}_{qs}(x,t)$,
where $\textbf{s}_{qs}(x,t)$ denotes the quasi--static solution
Eq.~\eqref{sol1} and $\delta\textbf{s}_{qs}(x,t)$ the deviation
from the exact solution, denoted as $\textbf{s}(x,t)$. For
$\delta\textbf{s}_{qs}(x,t)$ we have inside a single layer the
equation
\begin{eqnarray}\label{DSDE}
\frac{\partial \delta\textbf{s}_{qs}}{\partial
t}=\frac{|e|}{m}\delta\textbf{s}_{qs}\times \textbf{B}+D \triangle
\delta\textbf{s}_{qs}+{}\nonumber\\
\mu(\textbf{E}\cdot\nabla)
\delta\textbf{s}_{qs}-\frac{\delta\textbf{s}_{qs}}{\tau}-\dot{\textbf{s}}_{qs}~.
\end{eqnarray}
The inhomogeneity is defined as
\begin{equation}\label{source}
\dot{\textbf{s}}_{qs}:=\frac{\partial\textbf{s}_{qs}}{\partial
j}\frac{d j}{dt}+\left(\frac{d
\textbf{m}}{dt}\cdot\nabla_{\textbf{m}}\right)\textbf{s}_{qs}~,
\end{equation}
and is the source for a non--vanishing $\delta\textbf{s}_{qs}$.
Let us discuss the spin--relaxation in the ferromagnetic layers.
Here the typical relaxation time is $\tau\approx 1$~ps and it is
reasonable to neglect, in a first approximation, the Larmor,
diffusion,  and drift terms. The Larmor term is of the order of
$1/\omega$, the diffusion is characterized by a time scale
$\tau_d=l^2/D=\tau(l/\lambda)^2$, where $l$ is a characteristic
finite length (layer thickness). The drift term goes as
$\tau_j=l/|v_d|$. For $l=3$~nm (analyzer thickness as a worst
case) this gives $\tau_d\approx 0.3\tau$ and $\tau_j\approx
0.03$~ns for $j\approx 10^8$~A/cm$^2$. If we integrate
Eq.~\eqref{DSDE} under this assumptions we find as a first--order
correction (for a ferromagnetic layer),
\begin{equation}\label{dsdt}
\delta\textbf{s}_{qs}^{(1)}(x,t)=-\int_0^t
dt'e^{-(t-t')/\tau}\dot{\textbf{s}}_{st}(x,t')~.
\end{equation}
If we consider now a spacer layer ($x_1\ldots x_2$ in
Fig.~\ref{G0}) and compare $\tau$ with $\tau_d=\tau(l/\lambda)^2$
using the parameters given in Tab.~I we can see that $\tau_d\ll
\tau$. Diffusion is dominant in the spacer--layers. It occurs on a
time--scale $\tau_d\approx 10^{-3}$~ps. In fact, this quite
different time--scales in different layers are the reason why an
integration of Eq.~\eqref{SDE} by a discretization--procedure used
in usual PDE--toolboxes leads to numerical problems. To obtain an
estimate of $\delta\textbf{s}_{qs}^{(1)}(x,t)$ inside the
spacer--layer we solve Eq.~\eqref{DSDE} with boundary--conditions
given by Eq.~\eqref{dsdt}. Numerical results of this strategy to
estimate the accuracy of the QSE will be given below.
\section{\label{sec4} Seven--layer system}
Fig.~\ref{G0} shows the seven--layer structure, for which we apply
the general formalism. We select this system because such
structures where used for low--critical current experiments
\cite{Fuchs2,Meng}.  This allows testing of the present approach.
As indicated in the figure we use two opposite aligned
polarizer--layers, polarizer $P_1$ points in the $+z$ and
polarizer $P_2$ in the $-z$ direction. $P_1$ defines the parallel
position of the analyzer A, where a small deviation from $P_1$ is
needed for a non--vanishing initial--spin torque. Both polarizers
have the same material and geometric properties. As a consequence
the dipole field Eq.~\eqref{Hdd} vanishes exactly at the position
of the analyzer A and the anisotropy field Eq.~\eqref{han}
produces two energetically equivalent stable positions (degenerate
two--level system). We expect and proof that, if a current pulse
$j(t)$ switches the magnetization from P$\rightarrow$AP, then
$-j(t)$ does the inverse operation, AP$\rightarrow$P.
\subsection{Numerical strategy}

All computations are done with the help of \textit{Mathematica}. 
We use the solution
Eq.\eqref{sol1}-\eqref{ssenk} to compute the time-- and
space--dependent spin density inside of each layer for given
direction of $\textbf{B}$ and current $j$. The solution of the
total system requires the determination of all integration
constants $c_1\ldots c_{36}$. Whereas the boundary conditions
(continuous spin density and spin current density) are formulated
analytically, the solution is computed numerically as a function
of $\textbf{m}$ and $j$. The spin current density and the spin
torque in Eq.~\eqref{LLG} are then calculated self-- consistently
using Eq.\eqref{T}. The last step requires the numerical solution
of the LLGE, Eq.~\eqref{LLG}. 
\subsection{Optimized switching procedure}
We now address the switching of the analyzer magnetization (for
optimized switching using external magnetic fields see
\cite{Sun}). We first note that, due to the non--linearity in
$\textbf{m}$ of the LLGE, it is impossible to identify a single
current pulse profile which switches both from P$\rightarrow$ AP
and AP$\rightarrow$ P (initial--state--independent switching).
However, using the symmetry of the structure, one can identify a
current pulse profile which, when changing the current direction
only, promotes
both processes.\\
To find a simple pulse--shape which performs the desired task it
is convenient to use an optimization procedure based on a suitably defined
cost--functional $J$ \cite{Roloff}. We set
\begin{equation}\label{cost}
J=\|\textbf{m}(t_f)-\textbf{m}_T\|~,\hspace{5mm}0\leq J\leq 2~,
\end{equation}
where $\textbf{m}_T$ is the target magnetization and
$\textbf{m}(t_f)$ is its actual value at the prescribed target
time $t_f$. We choose the time--dependent current as
\begin{eqnarray}\label{curropt}
j(t;X_1,X_2,X_3)={}\nonumber\\
X_A \exp\left\{-\frac{X_B}{t_f^2}(t-t_f/2)^2\right\}+\sum_{l=1}^3 X_l
\sin(l \pi t/t_f)~,
\end{eqnarray}
with 3 variational parameters $X_1,\ldots X_3$ (one can
use also more parameters. Global optimization algorithms however
work best with a few parameters). The Gauss--pulse, characterized
by $X_A$, $X_B$, is selected by hand such that it is sufficient to
switch the magnetization from P$\rightarrow$AP. It is used as a
reference pulse.  However,  to steer the magnetization in the
prescribed time additional current contributions are needed. The
additional terms in Eq.~\eqref{curropt} are constructed to ensure
that at the end points of the control--time interval $[0,t_f]$ the
current vanishes for arbitrary $X_{1,2,3}$. To find the minimum of
$J(X_1,\ldots X_3)$ a standard line search method or genetic
algorithm can be used.
\subsection{System--Parameters}
We use the material--parameters typical for a Cu$_{\infty}$
/Fe$_{15}$/Cu$_3$/Py$_2$/Cu$_3$/Fe$_{15}$/Cu$_{\infty}$ (in nm)
multilayer--system. The relevant material--parameters are listed
in Tab.~I \cite{Stiles3,Reilly,Ziese}. We use a
material--independent electrical conductivity and
free--electron--density of $n=84/$nm$^3$.
\begin{table}
\begin{tabular}{|c|c|c|c|c|c|}
  \hline
  mat. & $\lambda$~ [nm]&  $\tau$~ [ns] &  $M_s$~ [A/m] & $\omega$ [GHz] & $\mathcal{P}$ \\
  \hline
Cu  & 450 & 0.024 & 0 & 0 & 0\\
Fe & 5 & 0.001 & $17\times 10^5$ & 230 & 0.45\\
Py & 5 & 0.001 & $8\times 10^5$ &110 & 0.37\\
  \hline
 \end{tabular}
 \caption{Material--parameters used in the simulation.}
 \end{table}
We obtain a microscopic expression for the coupling constant $\xi$
given by $\xi=-\frac{|e|n\hbar}{2mM_s d}=-0.969/d$, where
$d=x_3-x_2$ is the analyzer thickness.  For the Gilbert damping
parameter we set $\alpha=0.01$ \cite{Fuchs}. The direction of the
anisotropy field is chosen as
$\textbf{n}=(0,\sin(\varphi),-\cos(\varphi))$, with
$\varphi=0.9\pi$, and its modulus $\omega_{an}=2$~GHz.
\subsection{Results}
\subsubsection{Switching into constant current}
\begin{figure}
 \begin{center}
 \includegraphics[height=85 mm, angle=0]{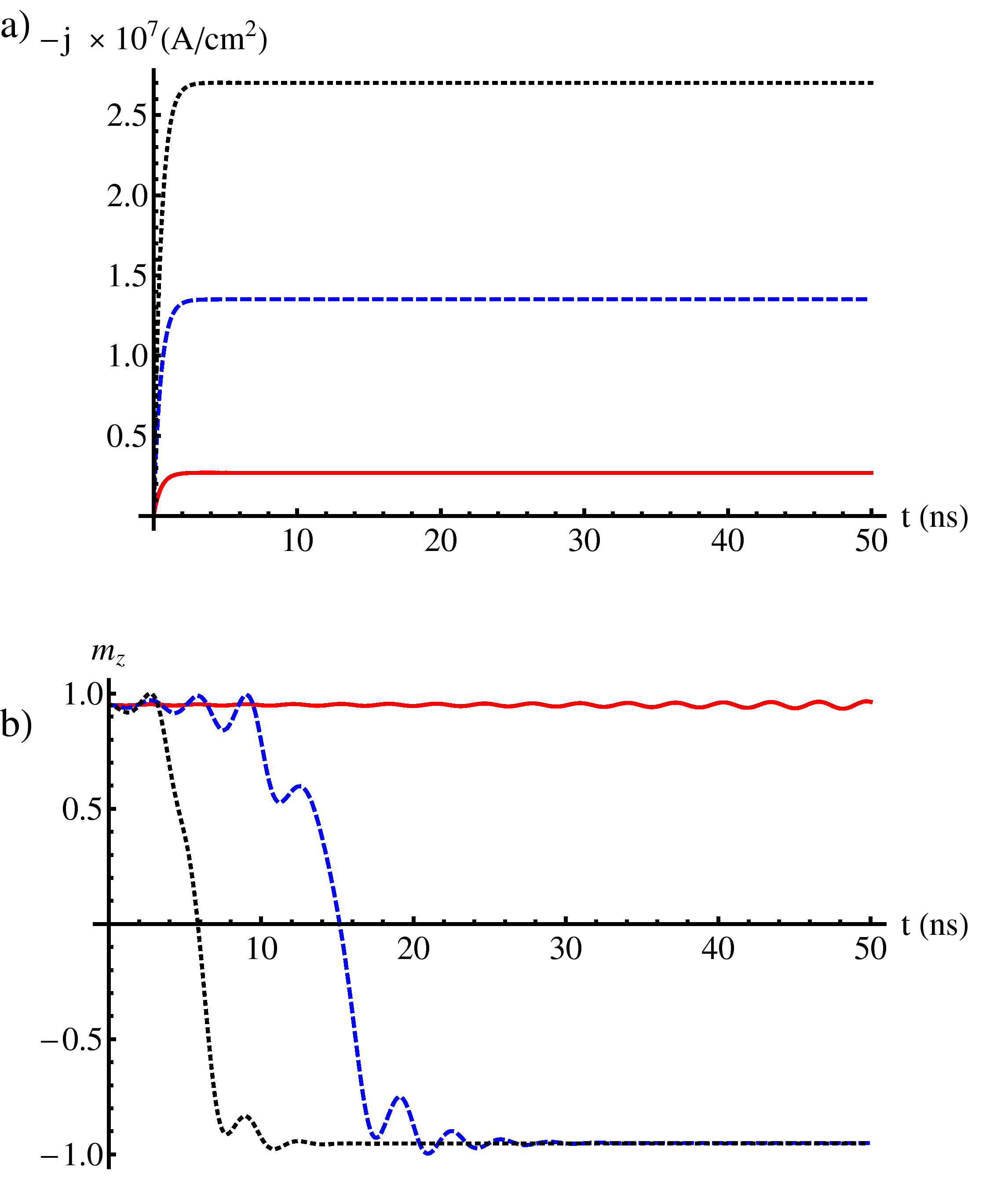}
 \caption{\label{G001}(color online) a) Electric current and b) $z$--component
 of the magnetization in the analyzer versus time.
Associated quantities are plotted in the same line style. For
decreasing current the switching time increases. The electric current is
plotted as $-j$ according to a positive drift velocity.}
 \end{center}
 \end{figure}

For a first example we consider the dynamics of the seven--layer
system in Fig.~\ref{G0} for a current that we switch on according
to $j(t)=j_0 (1-e^{-t/T})$, with $T=0.5$~ns. We have integrated
the LLGE for different values $j_0$, as shown in part a) of
Fig.~\ref{G001}. Part b) shows the  $z-$component of the
magnetization as a function of time. In all three cases the
magnetization switches from P$\rightarrow$AP, however, the lowest
current leads to a switching time of more than 100~ns. These
investigations agree well with basic experimental results in the
literature \cite{Fuchs2,Meng}: the critical current $|j_c|$ is of
the order $|j_c|\approx 10^6$~A/cm$^2$, for the parameters chosen
here, and depends on the saturation magnetization, Gilbert
damping, and anisotropy field \cite{Stiles}. As seen in
Fig.~\ref{G001}, switching into a constant spin--polarized current
leads to damped oscillations of the magnetization vector. Above
the critical current, they result in a flipping of the
magnetization vector into the new (AP) equilibrium position. For
currents $|j_0|<|j_c|$ one induces damped oscillations without
switching.  We should remark that the equilibrium--positions of
$\textbf{m}$ for a constant (spin--) current are no more given by
the directions of~ $\pm\textbf{n}$, but there is small deviation
due to the spin current, however, not resolved in Fig.~\ref{G001}.

\begin{figure*}
 \begin{center}
 \includegraphics[height=160 mm, angle=-90]{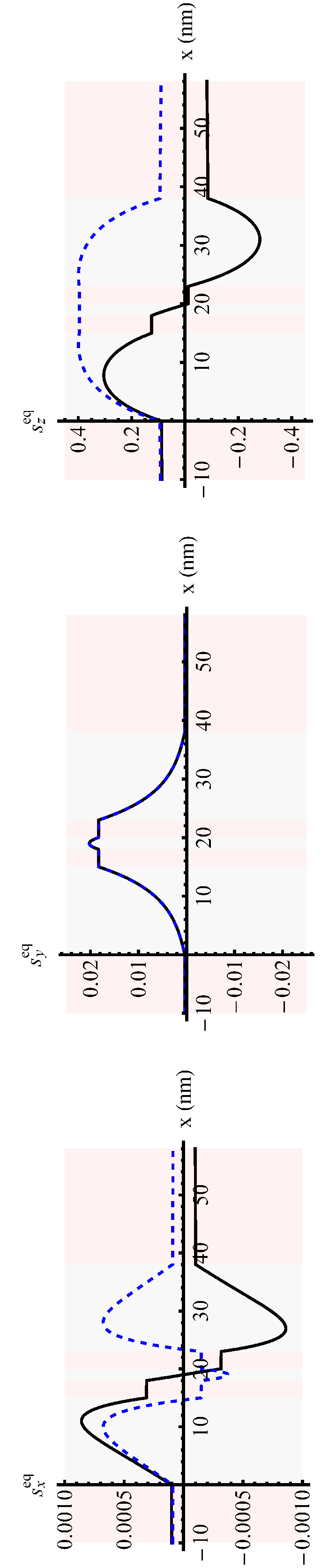}
 \caption{\label{G111}(color online) Equilibrium spin density
for two different polarizer alignments.
 The blue--dashed line corresponds to parallel
 orientations of the two polarizers, the black--solid line is for
 anti--parallel orientation, as used in low--current spin--torque experiments.
 The analyzer is at the position $\textbf{m}=\textbf{n}$ and therefore small
 perpendicular components of the spin density are present.}
 \end{center}
 \end{figure*}

The seven--layer structure with antiparallel
polarizer--orientations is crucial for the occurrence of low
$|j_c|$. Computations for parallel polarizer orientations ($P_1 \|
P_2$) give vastly different critical currents for P$\rightarrow$
AP and AP$\rightarrow$P flips. Fig.~\ref{G111} reveals the reason
for this result. For anti--parallel orientation of the polarizers
the $z-$component of the spin density shows a large gradient
inside the analyzer layer. As a consequence large spin currents
can be generated compared to parallel oriented polarizers. In fact
for a simplified model with vanishing dipole field (for sample
radius $R\rightarrow\infty$) and parallel polarizers the critical
current is $|j_c|>10^8$~A/cm$^2$ for this structure. As
investigated, switching times for the analyzer magnetization tend
to decrease with increasing $|j_0|$ \cite{Koch}.
\subsubsection{Optimal pulse--sequences}
We now consider the problem of switching of the magnetization
$\textbf{m}$ using an optimized time--dependent electric current,
where we set the switching time to $t_f=5$~ns. The first current
pulse should switch the magnetization from P$\rightarrow$AP.
Initial and desired final value of the analyzer magnetization
$\textbf{m}(t)$, respectively,  are
\begin{equation}
\textbf{m}(0)=\textbf{n}~\mbox{and}~
\textbf{m}(t_f)\stackrel{!}{=}\textbf{m}_T=-\textbf{n}~.
\end{equation}
A numerical minimization of Eq.~\eqref{cost}, limiting ourselves
to the pulse shape Eq. \ref{curropt},  gives as a result the first
pulse shown in Fig.~\ref{G2}. We stopped the computation when the
cost functional was $J\approx 0.006$. This means that the optimal
control pulse, rather than relying on intrinsic Gilbert damping,
actively drives the magnetization precisely  into the target state
AP; likewise for the back flip, see Fig.~\ref{G2} (c).   Note that
the pulse shape is chosen such that the current is zero at the
boundaries of the time interval. To ensure that the magnetization
remains in the AP state after the first flip, a few ns later we
apply the same pulse once more. Only a weak deviation from the
equilibrium position in form of a few damped oscillations are
visible demonstrating stability, see Fig.~\ref{G2}. However when
we apply the same pulse profile with opposite current direction we
switch the magnetization back from AP$\rightarrow$P. In addition
to $\textbf{m}(t)$ we have plotted in Fig.~\ref{G2} the
time--dependent spin density during the first current pulse. The
rows (a) and (b), respectively,  show the equilibrium spin density
and its deviation from equilibrium inside the multilayer device.
One observes the degree to which the equilibrium spin density
depends on the time--dependent magnetization $\textbf{m}(t)$:  due
to the choice of the magnetization of $P_1$ and $P_2$ (as
collinear) only the $z$--component of the spin density shows
significant deviation from equilibrium. The order of magnitude of
the deviation of $x$-- and $y$--components is of the order of the
error made by the QSE. The non--equilibrium spin density as
function of time is influenced by the actual position of
$\textbf{m}(t)$ and the current $j(t)$, as well as the
magnetization of $P_1$ and $P_2$ .
 \begin{figure*}
 \begin{center}
 \includegraphics[height=140 mm, angle=-90]{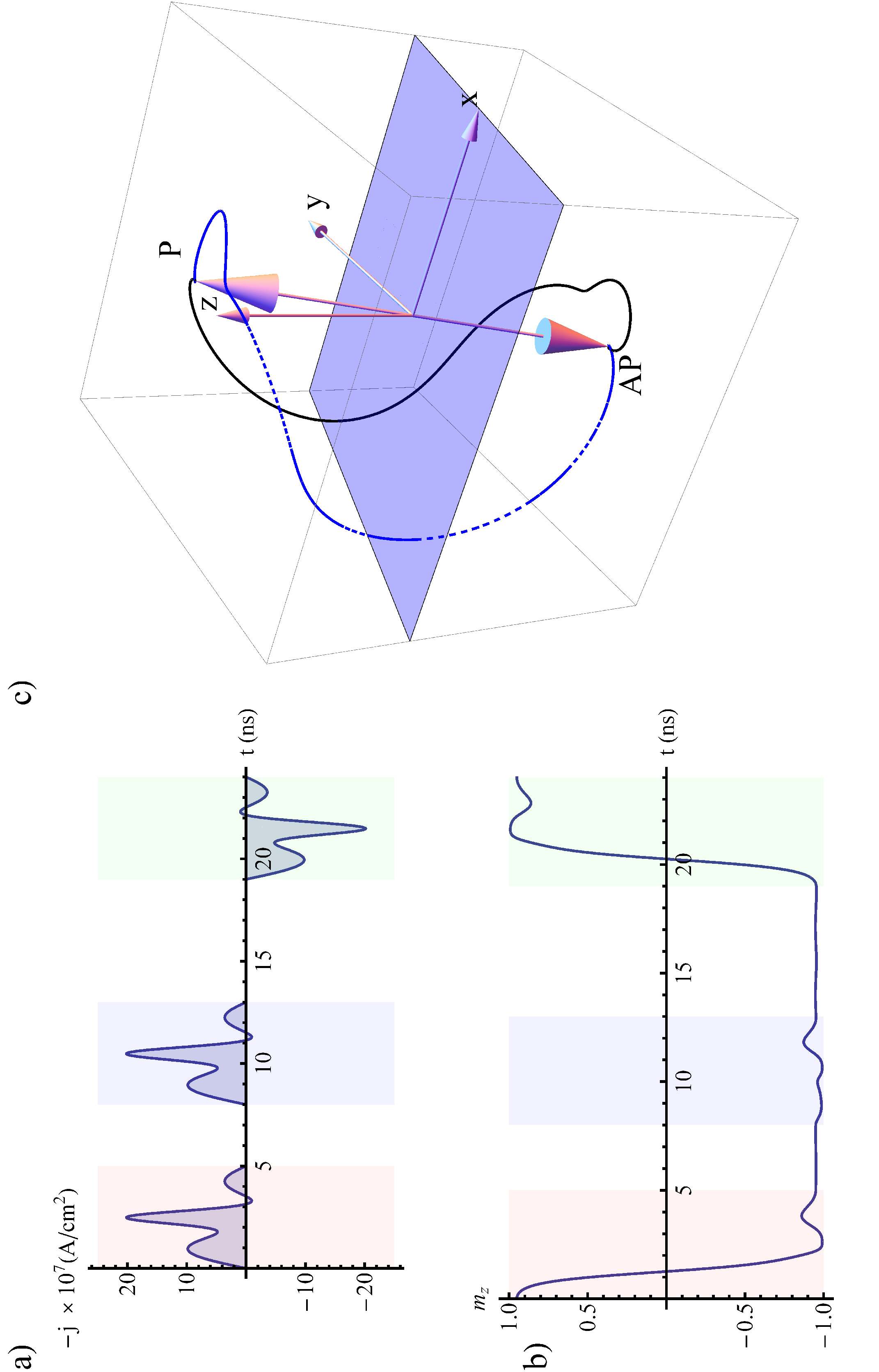}
 \caption{\label{G1}(color online) a) Optimized time--dependent electric current for switching
the analyzer magnetization from P$\rightarrow$AP and vice versa.
Three 5~ns pulses with the same shape are applied. The first
switches from P$\rightarrow$AP, the second is used to test
stability, and the third switches back AP$\rightarrow$P. b)
$z$--component of the analyzer magnetization vector
$\textbf{m}(t)$ as a function of time. c) Plot of the
three--dimensional trajectory of $\textbf{m}(t)$ during the first
(black, solid) and last pulse (blue, dashed).}
 \end{center}
 \end{figure*}
\subsubsection{Error estimate}
\begin{figure}
\begin{center}
\includegraphics[height=80 mm, angle=-90]{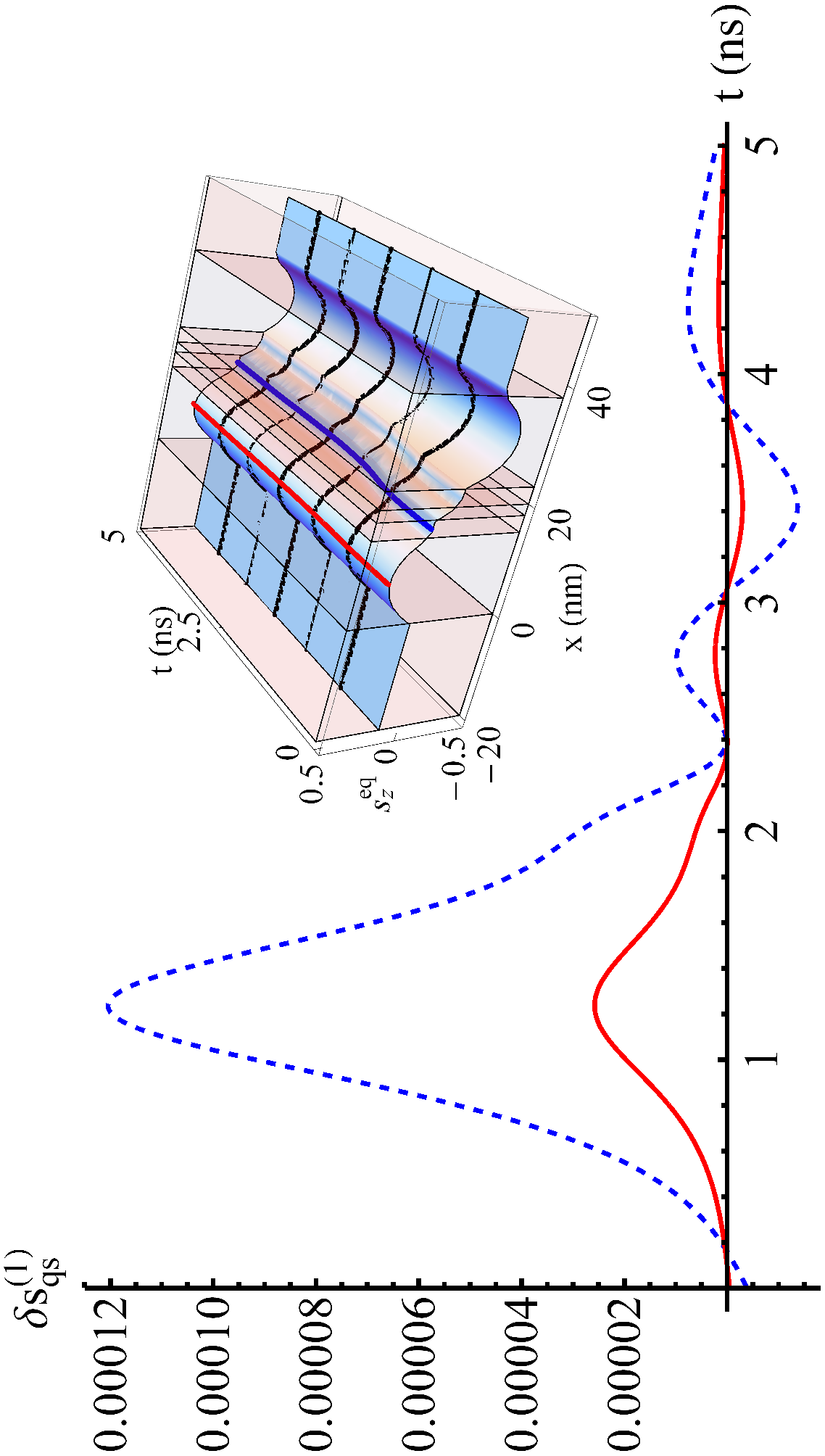}
\caption{\label{Gcomp}(color online) Numerical estimate of the
error within the QSE relative to an exact treatment of the SDDE. The
inset shows the locations where we compute
$\delta\textbf{s}_{qs}^{(1)}$. Inside the polarizer (red--solid line)
we use Eq.~\eqref{dsdt} to estimate the deviation from the exact result, whereas inside the spacer--layer,
Eq.~\eqref{DSDE} is integrated numerically.}
\end{center}
\end{figure}

We have used the results  from the previous section to test the
numerical validity of the stationary solution as discussed in
Sec.~\ref{Fulltimedependent}. Fig.~\ref{Gcomp} shows the estimate
for the deviation of the  $z-$ component of the spin density in
selected parts of the structure. The solid line is for the center
of $P_1$, while the dashed line is for the center of the spacer
layer to the left of the analyzer. The figure shows that $[\delta
\textbf{s}_{qs}^{(1)}]_z$, depending on position, is of the order
of $10^{-5}-10^{-4}$, compared with $s_z^{eq}\approx 0.1-0.3$ (see
Fig.~\ref{G111}). The dominant contribution in Eq.~\eqref{source}
comes from the moving magnetization, whereas the current
contribution is neglibile. For the other components we obtained
similar results regarding relative errors.
\begin{figure*}
 \begin{center}
 \includegraphics[height=170 mm, angle=-90]{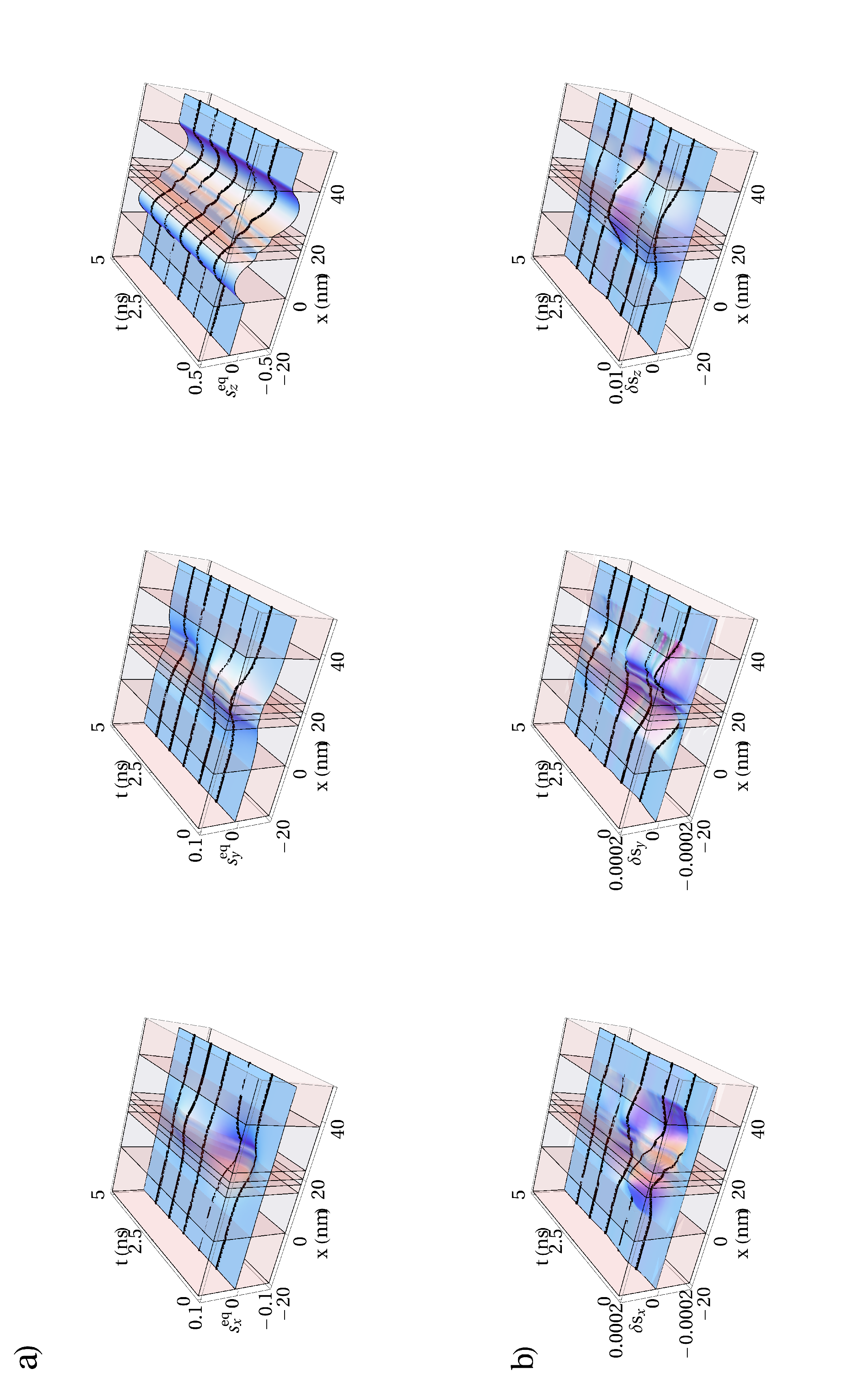}
 \caption{\label{G2}(color online) a) The first row shows the
 equilibrium spin density for the switching process
 P$\rightarrow$
 AP for the first pulse in Fig.~\ref{G1}. It depends on $\textbf{m}(t)$.
b) Non--equilibrium spin density induced by the
electrical--current pulse. The computed values for $\delta s_x$,
$\delta s_y$, however, are at the limit of the accuracy of the
QSE.}
 \end{center}
 \end{figure*}
\section{\label{sec5} Conclusions and outlook}
We have presented a self--consistent model for magnetization
switching by spin--polarized electric current in metallic
ferromagnetic heterostructures. Our method is founded upon an
analytic solution of the stationary spin drift--diffusion equation
(SDDE) for each layer using constant material parameters, electric
current, and magnetic field. Matching layers, using continuity of
spin density and spin current density at the interfaces as
boundary conditions, we obtain an analytic solution for the spin
density of the entire heterostructure. Making a quasi--static
approximation in which the time dependence of the spin density
depends on time solely via the electric current and net magnetic
field, the time evolution of the spin density is computed in
parallel to the Landau--Lifshitz--Gilbert (LLGE) equation. Both
equations couple via the spin torque effect and the
time--dependent magnetization in the SDDE. This method allows for
an efficient and robust mathematical description of the coupled
carrier spin and magnetization dynamics in metal/ferromagnet
heterostructures. Because the model is based on a completely
analytic solution of the stationary SDDE for given electric
current and magnetic field for each layer, it is applicable to
heterostructures of high complexity, for example for  tilted
polarizers or structures exposed to external magnetic fields
\cite{He}.\\
We have demonstrated the efficiency of this semi--analytic
approach by investigating a seven--layer system with antiparallel
oriented polarizers, as studied in recent experiments, and
computed optimized current pulses to switch the magnetization from
P$\rightarrow$AP$\rightarrow$P in specified time of 5~ns. As
expected for the system under investigation, the obtained current
densities are in the range of $10^8$~A/cm$^2$, with a critical
current of about  $10^6$~A/cm$^2$. Using optimal control theory,
we identify solutions for current profiles which allow for precise
switching in predetermined switching times. We provide and discuss
one example.\\
Furthermore, a detailed investigation of the validity of the
quasi--static time evolution of the SDDE is given. It confirms
excellent accuracy for the example of the simulated seven--layer
heterostructure.\\
Several future applications of the presented formalism can be
envisioned. A combined variation of material-- and geometric
parameters to obtain optimal current pulses with low critical
currents. A description of thermal fluctuations using
temperature--dependent effective (Langevin--) fields in the LLGE
(via the spin torque in the SDDE) and the search of ''thermally
robust'' current pulses by averaging over many field
configurations.

 \begin{acknowledgments}
We wish to acknowledge financial support of this work by FWF
Austria, project number P21289-N16.
 \end{acknowledgments}
%
\appendix
\section{\label{AppA} Stationary solution of the SDDE}
Here we summarize the remaining analytic expressions for the stationary
solution and constant material parameters as presented in
Sec.~\ref{SDDstatsol}. We use the dimensionless quantities
$\kappa:=\omega\tau$ and $\rho:=l_d/\lambda$. Further we define
\begin{eqnarray}
a=\mathrm{Re}[\sqrt{4+\rho^2+4i\kappa}]={}\nonumber\\
\sqrt{\frac{4+\rho^2}{8}+
\sqrt{\frac{\kappa}{2}\left[1+\left(\frac{4+\rho^2}{4\kappa}\right)^2\right]}}~,
\end{eqnarray}
\begin{eqnarray}
b=\mathrm{Im}[\sqrt{4+\rho^2+4i\kappa}]={}\nonumber\\
\sqrt{-\frac{4+\rho^2}{8}+
\sqrt{\frac{\kappa}{2}\left[1+\left(\frac{4+\rho^2}{4\kappa}\right)^2\right]}}~.
\end{eqnarray}
Using the auxiliary functions,
\begin{equation}
F_1(x)=e^{-\frac{\rho x}{2
\lambda}}\cos\left(\frac{bx}{2\lambda}\right)\sinh\left(\frac{ax}{2\lambda}\right)~,
\end{equation}
\begin{equation}
F_2(x)=e^{-\frac{\rho x}{2
\lambda}}\cosh\left(\frac{ax}{2\lambda}\right)\sin\left(\frac{bx}{2\lambda}\right)~,
\end{equation}
\begin{eqnarray}
F_3(x)=e^{-\frac{\rho x}{2
\lambda}}\cos\left(\frac{bx}{2\lambda}\right)\cosh\left(\frac{ax}{2\lambda}\right)~,
\end{eqnarray}
\begin{eqnarray}
F_4(x)=e^{-\frac{\rho x}{2
\lambda}}\sin\left(\frac{bx}{2\lambda}\right)\sinh\left(\frac{ax}{2\lambda}\right)~,
\end{eqnarray}
the four dimensionless functions $G_i(x)$, entering in
Eq.~\eqref{ssenk} are:
\begin{widetext}
\begin{eqnarray}
G_1(x)=\left[-4a^3\rho\kappa+4 a \rho \kappa(4+3
b^2+\rho^2)\right]F_1(x)+{}\nonumber\\
\left[4 b^3 \rho\kappa+4 b \rho \kappa(4-3a^2+\rho^2)\right]F_2(x)-{}\nonumber\\
4\kappa(a^2+b^2)(-4+a^2-b^2-\rho^2)F_3(x)-{}\nonumber\\
8a b (a^2+b^2)\kappa F_4(x)~,
\end{eqnarray}
\begin{eqnarray}
G_2(x)=a \rho\left[a^4+(4+5b^2+\rho^2)(4-2a^2+b^2+\rho^2) \right]F_1(x)+{}\nonumber\\
b \rho\left[5a^4+(4+b^2+\rho^2)^2-2a^2(12+5b^2+3\rho^2)\right]F_2(x)-{}\nonumber\\
(a^2-b^2)\left[(a^2+b^2-\rho^2-4)^2-4a^2b^2\right]F_3(x)+{}\nonumber\\
4a b (a^2-b^2)\kappa\left[4-a^2+b^2+\rho^2\right] F_4(x)~,
\end{eqnarray}
\begin{eqnarray}
G_3(x)=\left[24a^2b\kappa-8b\kappa(4+b^2+\rho^2)\right]F_1(x)+{}\nonumber\\
\left[24ab^2\kappa+8a\kappa(4-a^2+\rho^2)\right]F_2(x)~,
\end{eqnarray}
\begin{eqnarray}
G_4(x)=\left[-8a^3\kappa+8a\kappa(4+3b^2+\rho^2)\right]F_1(x)+{}\nonumber\\
\left[8b^3\kappa+8b\kappa(4-3a^2+\rho^2)\right]F_2(x)~.
\end{eqnarray}
\end{widetext}
The integration of the normal component of Eq.~\eqref{SDEstat}
requires the solution of two second order differential equations.
It is advantageous to transform this two second order equations
into four first order equations and solve this system by matrix
exponentiation. This procedure, after some simplifications, leads
to the four functions $G_{i}(x)$, which build the fundamental
solution.


\end{document}